\documentclass[conference]{IEEEtran}
\makeatletter
\def\ps@headings{%
\def\@oddhead{\mbox{}\scriptsize\rightmark \hfil \thepage}%
\def\@evenhead{\scriptsize\thepage \hfil \leftmark\mbox{}}%
\def\@oddfoot{}%
\def\@evenfoot{}}
\makeatother
\ifCLASSINFOpdf
 \else
\fi

\usepackage[cmex10]{amsmath}
\usepackage{amssymb}
\usepackage{algorithmic}

\usepackage{algorithm}
\usepackage{array}
\usepackage{mdwmath}
\usepackage{mdwtab}
\usepackage{mathrsfs}
\usepackage[belowskip=-15pt,aboveskip=0pt]{caption}
\usepackage{epstopdf}
\usepackage{graphicx}
\usepackage{subfigure}
\usepackage{verbatim}

\DeclareGraphicsExtensions{.eps}
\usepackage{indentfirst}
\usepackage{url}
\newtheorem{lemma}{\textbf{Lemma}}
\newtheorem{theorem}{\textbf{Theorem}}

\newenvironment{sequation}{\begin{equation}}{\end{equation}}

\newcommand{\eg}{{\it e.g.}}
\newcommand{\etal}{{\it et al.~}}
   
\newcommand{\ie}{{\it i.e.}}

\newenvironment{icompact}{
  \begin{list}{$\bullet$}{
    \parsep 1pt plus 1pt
    \partopsep 1pt plus 1pt
    \topsep 1pt plus 2pt minus 1pt
    \itemsep 1.5pt plus 1pt
    \parskip 0pt plus 2pt
    \leftmargin 0.15in}
       }
  {\normalsize\end{list}}

\begin{document}

\thispagestyle{plain}
\setlength{\belowcaptionskip}{-3pt}
%
\title{A Collaborative Framework for In-network Video Caching in Mobile Networks}

\author{}
\author{
\IEEEauthorblockN{Jun He\IEEEauthorrefmark{1}, Honghai Zhang\IEEEauthorrefmark{2}, Baohua Zhao\IEEEauthorrefmark{1}, Sampath Rangarajan\IEEEauthorrefmark{2}}
\IEEEauthorblockA{\IEEEauthorrefmark{1}School of Computer Science and Technology, University of Science and Technology of China}
\IEEEauthorblockA{\IEEEauthorrefmark{2}NEC Laboratories America}
\IEEEauthorblockA{Email: myname@mail.ustc.edu.cn, bhzhao@ustc.edu.cn, \{honghai,sampath\}@nec-labs.com}
}

\maketitle

%
\IEEEpeerreviewmaketitle

\begin{abstract}
Due to explosive growth of online video content in mobile wireless networks,
in-network caching is becoming increasingly important to improve the end-user
experience and reduce the Internet access cost for mobile network operators.
However, caching is a difficult problem
due to the very large number of online videos and video  requests,
limited capacity of caching nodes,
and limited bandwidth of in-network links.
Existing solutions that rely on static configurations
and average request arrival rates
are insufficient to  handle
dynamic request patterns effectively.
In this paper, we propose a dynamic collaborative video caching framework to be
deployed in mobile networks.
We decompose the caching problem into a content placement subproblem
and a source-selection subproblem.
We then develop SRS (System capacity Reservation Strategy) to solve the content
placement subproblem, and  LinkShare, an adaptive traffic-aware algorithm
to solve the source selection subproblem.
Our framework supports congestion avoidance
and allows merging multiple requests for the same video into one request.
We carry extensive simulations to validate the proposed schemes.
Simulation results show that our SRS algorithm achieves performance
within $1-3\%$ of the optimal values
and LinkShare significantly outperforms existing solutions.
\end{abstract}

\section{Introduction}\label{sec:introduction}
 Video content distribution and  caching
have been studied extensively in the past two decades
\cite{Borst2010}\cite{Michel1998}\cite{Rodriguez2001} because they
 can effectively reduce the end-to-end delay and network traffic.
Recent years have witnessed an explosive growth of video delivery
over mobile wide-area wireless data networks (\eg, LTE) \cite{Dahlman2008}
due to the proliferation of smart phones and tablets.
Video content caching faces new challenges attributed to the huge number of
online videos (in the order of hundreds of millions),
very high video rates, and limited
storage sizes and network bandwidth.
As a result, it has received revived
interest  recently\cite{applegate2010optimal,haiyongxie2012tecc}.

\begin{figure}[t]
\setlength{\abovecaptionskip}{5pt}
\setlength{\belowcaptionskip}{-10pt}
\centering
\includegraphics[width=0.95\linewidth, height=2.0in]{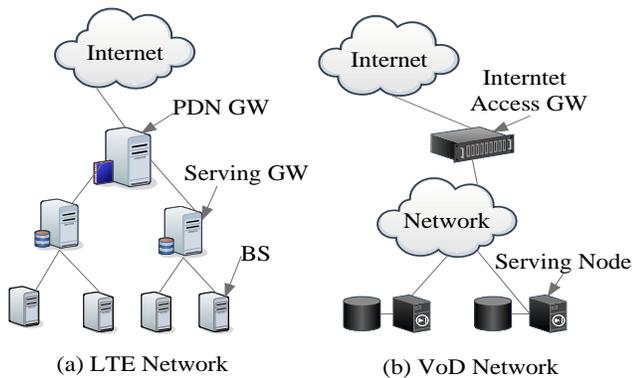}
\vspace{-0.1in}
\caption{A basic LTE serving network and a generalized video caching model.}
\vspace{+0.1in}
\label{fig:lte}
\end{figure}

In this work, we consider the video caching problem in mobile networks
where the caching nodes are distributed along with the mobile gateways.
As an example, Fig. \ref{fig:lte}(a) shows a basic LTE mobile core network.
A PDN (Packet Data Network)  gateway
 provides connectivity to
the external Internet and connects Serving gateways (S-GWs) internally.
Each serving gateway connects a set of base stations,
which offer wireless service to the user equipments (UEs).
If 
the mobile core network does not employ video content caching,
it simply relays the video requests made from the users and
fetches the data from the external Internet.
Typically, mobile network operators and the Internet service providers are
different entities.
As a result, requesting data from the Internet not only incurs
extra delay but also introduces higher Internet access costs for the mobile network operators.
Therefore, deploying content caching service in  a mobile core network
improves the end-user experience and simultaneously
reduces the OPEX (Operational Expense) for the mobile network operators.
A generalized caching system model in a mobile network is described in
Fig. \ref{fig:lte}(b).

To address the challenge of delivering a huge number of video clips
within the current mobile network architecture,
we consider collaborative distributed caching,
where the caching nodes are co-located with the Serving gateways.
In such systems, multiple caching nodes jointly  cache all videos that are of
interest and each  of them simultaneously attempts to maximize the cache hit ratio
of the clients in its own domain.
With collaborative caching, when a request arrives at a serving gateway,
it first checks whether the video is cached in its local cache.
If yes, the cached video clip is delivered directly to the requesting client.
Otherwise, it looks for the video (possibly through a directory service)
from other in-network caching nodes.
If no copy is found in the system, the request is relayed to the external
Internet via the PDN gateway.  Optionally, the PDN gateway may also host a
caching server.

In our collaborative caching framework,
we aim to minimize the aggregate cost of data transfer in the network
subject to the storage capacity limit and link bandwidth constraints.
The cost of data transfer is defined as the sum of cost on all links,
which is defined as  a convex function of the link loading (to model transmit
costs).
Our framework addresses two important problems.
(1) How to place all videos among the caching nodes (content placement
problem)?
(2) Which caching nodes are selected to fetch the requested video
(source selection problem)?

In contrast to existing solutions in
\cite{applegate2010optimal, haiyongxie2012tecc}
that solve the joint content placement and source selection problem,
we consider these two problems separately
because we believe that they  should be
solved at different time scales.
It is hard to move all video content, and it may take a long time
to even find a solution for the content
placement problem (e.g., it takes more than one hour to even find a sub-optimal
solution in \cite{applegate2010optimal}).
Thus, the content placement problem should be solved over a long period of time.
On the contrary, the source selection problem should be solved
instantaneously on each caching node to respond to rapid change of
traffic arrival patterns and dynamic network conditions.
Therefore, we solve these two problems  independently.
We solve the content placement problem with the aim of maximizing overall
cache hit ratio while ensuring all video clips are cached in the system.
For the source selection problem, we divide time into rounds and
route in-network requests dynamically in each round to respond to
instantaneous request patterns and link states.
By decoupling the two problems, our proposed schemes  are
more practical and more efficient.

We make three important contributions in this work.
\noindent
Firstly, we propose a complete framework to solve the
in-network video caching problem.
Secondly,
we develop an efficient algorithm for the content placement subproblem
and a dynamic routing scheme, LinkShare, for
the source-selection subproblem.
In contrast to existing algorithms for the source selection problem
that typically rely on time-averaged video request patterns,
the LinkShare scheme is traffic-aware and fine-grained,
and considers {\em instantaneous} video request patterns and link state
information.
Thirdly, we show that our proposed schemes also support
instantaneous network congestion avoidance and can
merge multiple requests for the same video around the same time into one
request.

We perform extensive simulations to validate the proposed schemes.
Our simulation results indicate that our framework provides
an efficient solution to the in-network caching problem and
is more robust under burst request patterns.

The rest of the paper is organized as follows. Section II presents the related work.
Section III describes the system model.
Section IV and section V present the proposed solutions
to the problems described in section III.
Simulation results are presented in section VI. Section VII concludes the paper.

\section{Related Works}\label{sec:rw}

Online content placement and replication have attracted extensive attention.
For general content distribution problems,
we refer readers to the survey by Androutsellis and
Spinellis \cite{Androutsellis-Theotokis2004} and the references therein.

Several recent works have considered a joint design of collaborative caching and
routing (source selection).
Borst \etal \cite{Borst2010} proposed  
a caching scheme over hierarchical caching clusters with
the aim of maximizing the traffic volume served by caches as well as minimizing the total bandwidth costs.
However, they only developed solutions for the symmetric scenarios where the
request pattern is uniform across all caching nodes.

Applegate \etal\cite{applegate2010optimal}
formulated a MIP (Mixed Integer Programming) model to minimize the cost of the total data transfer,
subject to the disk space and link bandwidth constraints.
However, the presented solution therein is of very high complexity.
Even though efficient algorithms such as the potential function method
\cite{bienstock2002potential} were employed,
it still took more than one hour to find $\epsilon$-suboptimal solutions even
for a relaxed LP (linear programming) version of the problem.
Moreover, the work \cite{applegate2010optimal} assumed
long-term average request pattern in their problem formulation
and thus did not consider the burstiness of the user requests.

Xie \etal\cite{haiyongxie2012tecc} considered a joint traffic engineering
and collaborative caching problem over an unstructured flat network model
with the objective of minimizing maximum congestion level from ISPs' perspective.
By contrast, we assume a convex cost function and show that finding a feasible
solution to our problem is equivalent to the problem studied in
\cite{haiyongxie2012tecc}.

The source selection subproblem is similar to the multi-commodity flow problem
\cite{garg2007,Ouorou2000}.
Jiang \etal \cite{jiang2009cooperative} and
DiPalantino \etal \cite{dipalantino2009traffic}
studied the source selection problem for both the objectives of traffic
engineering and content distributions.
They developed algorithms based on game theory.
Our source selection algorithms differ from the above references in that
we deal with a system with continuously changing request patterns.
We address the challenge of rapid fluctuation of the request patterns,
and design dynamic solutions corresponding to instantaneous link states.
%

\section{System Model}
We consider a caching system in an LTE mobile core network as depicted in Fig.
\ref{fig:lte}, where every serving gateway has a local
cache holding a subset of all available video clips,
and the serving gateways are inter-connected via the PDN gateway and
possibly other network routers.
A serving gateway receives and satisfies all video requests from users associated
with the base stations it serves.
If a requested video is at the local cache, the local copy is fed to the
clients.
Otherwise, the serving gateway
determines, possibly through a directory service,
whether any other serving gateways
have a copy of the data.
If so,
the serving gateway will choose one of them to fetch the data
and serve the client's request. Otherwise, it passes the request to the PDN
gateway, which in turn sends the request to the original server through an
ISP network. A PDN gateway may also have its own cache to serve requests that
are not found in the serving gateways.

 \begin{table}[h!]
\centering
\caption{Basic Notaions} \label{table:notations}
\begin{tabular}[c]{|c|l|}
\hline
Notation & Meaning\\ \hline
$\mathcal{N}$  &  The set of videos cached in the system \\ \hline
$\mathcal{M}$ & The set of serving nodes with caches \\ \hline
$\mathcal{L}$ & The set of in-network links \\ \hline
$D_i$& The caching capacity in node $i$ \\ \hline
$S_i$& The video set cached in node $i$ \\ \hline
$y^k_i$& Indicator for caching video $k$ in node $i$ \\ \hline
$x^k_{ji}$& Fraction of video $k$ delivering from node $j$ to $i$ \\ \hline
$\lambda^k_i$ & Aggregate request frequency for video $k$ in  node $i$\\ \hline
$s_k$ & The size of video $k$ \\ \hline
$r_k$ & The video rate of video $k$ \\ \hline
$P(j,i)$ & The link path from node $j$ to $i$ \\ \hline
$C_l$ & Link capacity of link $l$ \\ \hline
$d_{j,i}$&The cost of transferring one unit data from node $j$ to $i$ \\ \hline
$R_i$ & Set of videos requested at node $i$ but not cached there \\ \hline
$T_k$ & Set of nodes containing video $k$ \\ \hline
\end{tabular}
\end{table}

We consider such a collaborative caching system with a set $\mathcal{M}$ of
Serving nodes (i.e., the Serving gateways or PDN gateways with caching capacity),
which is deployed to jointly cache a set $\mathcal{N}$ of videos.
A video clip $k \in \mathcal{N}$ has size $s_k$ and data rate $r_k$.
Serving node $i$ has caching capacity of $D_i$ and caches video
subset $S_i\subseteq\mathcal{N}$. The aggregate request frequency
at node $i$ for
video $k$ is $\lambda^k_i$, which can be calculated and predicted from
historical statistics. In fact, $\lambda^k_i$ represents the
popularity of video $k$ at node $i$.

We define the cost of transferring one-unit of  data from node $j$ to node $i$
as the end-to-end delay $d_{j,i} = \sum_{l \in P(j,i)} \zeta_l(f_l)$,
where $P(j,i)$ is the path from node $j$ to $i$,
$\zeta_l$ denotes the link delay and is modeled as
a convex, non-decreasing, and continuous function of the total load $f_l$
    on the link $l$.
    We use indicator  variable $y^k_i$ to denote whether
    video $k$ is cached at node $i$ and
    $x^k_{ji}$ to represent the fraction of video $k$ served from node $j$
    to node $i$ to fulfill the requests at node $i$.
    Table \ref{table:notations} summarizes important notations used in the paper.

%
Our objective is to minimize the total (or average)
end-to-end delay in the caching system, subject to the disk storage and link
bandwidth constraints.
The corresponding problem includes two subproblems:
(i) the {\bf content placement subproblem} (i.e., what videos are stored on each
serving node?) and (ii) {\bf source selection subproblem}
  (i.e., where to fetch a video from the system?).
It is tempting to solve the joint problem simultaneously,
as is done in  \cite{applegate2010optimal,haiyongxie2012tecc}.
However, we note that these two subproblems should be solved
  at different time scales.
  The cache placement subproblem should be
  solved over a long period of time (\eg, on a weekly basis),
  as  it involves moving a large amount of data across the network.
  On the contrary,
  the source selection decision can be updated frequently
  depending on dynamic traffic demand,
  which varies significantly over a short period of time (\eg, in minutes or
  even seconds).
  Therefore, in this work, we develop the problem formulation for these two subproblems
  separately.

\subsection{Content placement subproblem} \label{ssubsection:gp}
For this subproblem, our objective is to maximize the  total cache
hit ratio at each local serving node, weighted by the size of each video,
subject to the disk space and the content coverage constraints.
It is formulated as the {Maximum Hit Problem} ({MHP}):
\begin{eqnarray}
    &\max\ & \sum_{i\in\mathcal{M}}\sum_{k\in\mathcal{N}}s_k \lambda^k_i  y^k_i \label{eq:shmpp} \cr
  &\textrm{s. t.}& \sum_{k\in\mathcal{N}} y^k_i s_k \le D_i,  \forall  i\in\mathcal{M}
  \label{eq:limcp-c1} \\
\label{eq:limcp-c4}
&&   \sum_{i\in\mathcal{M}}y^k_i \ge 1, \forall k\in\mathcal{N}
 \\
 \label{eq:limcp-c7}
&& \textrm{var.}~~  y^k_i\in \{0,1\},\forall i\in\mathcal{M}, k\in\mathcal{N}.
\end{eqnarray}
The first constraint above represents the storage limit
at serving node $i$.
The second one indicates that
at least one copy of video $k \in \mathcal{N} $ has to be
cached  in the mobile core network.
It is non-trivial to solve this problem, as it can
be shown to be strongly NP-hard.
Therefore, there is no polynomial or pseudo-polynomial
algorithm for problem  MHP unless P = NP.

\begin{theorem}
\label{theorem:smhpp-np}
It is strongly NP-hard
to find an optimal solution to the problem {MHP}.
\end{theorem}
The proof is omitted due to space limit and can be found in \cite{junhe2012a}.

\subsection{Source Selection Subproblem}
\label{ssubsection:ssp}
For this subproblem, 
we divide the time into rounds with duration
$\Delta t$. 
Within a round, each serving node collects the requests from the clients.
At the end of the round, the system aggregates all requests  and determines
the source selection for all the requests made at the present round.
By merging the requests for the same video during a round,
the serving nodes
can potentially save the bandwidth requirement,
although it is at the cost of some scheduling delay, which is upper bounded by
$\Delta t$. Choosing a larger $\Delta t$ increases the opportunity for
merging requests but at the price of higher scheduling delay.

Now we only need to consider
the set $R_i$ of videos requested at node $i$ but not cached at it
during the current round.
Let $T_k$ be the set of nodes containing a copy of video $k$.
For each link $l\in \mathcal{L}$, let $f^{bg}_l$ be the background traffic rate,
$f^{re}_l$ be the rate of the remaining traffic starting from previous
rounds, and $f^{ss}_l$ be the traffic rate generated in the present round
by the source selection algorithm. Then,
\begin{eqnarray}
 f^{ss}_l=\sum_{i\in \mathcal{M}}\sum_{k\in R_i}\sum_{j\in T_k:  l \in P(j,i)}
 x^k_{ji}r_k, \forall l\in\mathcal{L}.
 \label{eq:fssl}
\end{eqnarray}
The total loading on link $l$ is
\begin{sequation}\label{eq:fl}
     f_l=f^{ss}_l+f^{bg}_l+f^{re}_l.
\end{sequation}
The cost (delay) of fetching one unit of data from node $j$ to $i$ is
$$
    d_{ji}=\sum_{l\in P(j,i)}\zeta_l(f_l),
$$
where $\zeta_l(\cdot)$ is the link delay function.

We formulate this as the {Minimum Round Cost Problem} ({MRCP}):
\begin{eqnarray}
\label{eq:mrcp}
	&\min& \sum_{i\in\mathcal{M}}\sum_{k\in R_i}\sum_{j\in T_k}
    d_{ji} r_k x^k_{ji} \\
\label{eq:linkcapacity}
 &\textrm{s.t.} &
   f_l\le C_l, \forall l\in\mathcal{L} \cr
\label{eq:reachable}
&&   \sum_{j\in T_k}x^k_{ji}=1, \forall i\in \mathcal{M}, k\in R_i \\
   &&  \textrm{var.}~~ x^k_{ji}\in[0,1], \forall k\in\mathcal{N}, i,j\in\mathcal{M} 
\label{eq:limcp-c6}.
\end{eqnarray}
The objective here is to minimize the sum of  weighted cost.
The first constraint comes from the link capacity constraint.
The second and third imply that each video can be picked from multiple sources.
Our formulation is different from that in \cite{applegate2010optimal}
in that the link cost here depends on the loading of that link,
while  the link cost in \cite{applegate2010optimal} is a constant.

\section{Solutions to MHP}\label{sec:alg:smhpp}

As it is NP-hard, {MHP} cannot be solved optimally
in polynomial time unless P=NP.
In this section, we propose an efficient heuristic algorithm to solve
the problem.  The basic idea is to
reserve a fraction $1-\alpha$ of the total storage capacity
for maintaining full coverage of all videos and to use the rest capacity
at each serving node to cache the most frequently requested videos.
This is motivated by the fact that
the popularity of the videos typically has a Zipf-like distribution as
discussed in Section \ref{sec:introduction},
which suggests that only a small number
of popular videos are very frequently requested \cite{Adamic2000}.

\subsection{$\alpha$-MHP algorithm}

Assume that the fraction $\alpha$ for caching all videos in $\mathcal{N}$
is given, our scheme consists of four steps, which are summarized in Alg. \ref{alg:heuristic_smhpp}.
In Alg. \ref{alg:heuristic_smhpp},
$\mathcal{N}_0$ is the union of all cached video set for the system at step 1,
 and $\mathcal{S}_i$ and $D_i'$  are
the cached video set and the remaining capacity on the node $i$,
respectively.
$H(\alpha)$ is the maximum objective value found.

\begin{algorithm}
\caption {${\alpha}$-{MHP} Algorithm}
\label{alg:heuristic_smhpp}
\begin{algorithmic}[0]
\STATE [Step 1]:
Solve the Reservation Packing problem($\mathcal{N}, \alpha, D_i, i \in \mathcal{M}$),
 output  $\{\mathcal{S}_i,i\in\mathcal{M} \}$  and
 $\mathcal{N}_0 = \cup_{i \in \mathcal{M}} \mathcal{S}_i $.
 \STATE
For $i\in\mathcal{M}$, $D_i'=D_i - \sum_{k \in \mathcal{S}_i} s_k$. \label{alg:line:preservation}
\STATE [Step 2]:  $\mathcal{N}_r=\mathcal{N}\setminus\mathcal{N}_0$.
\STATE Solve OCMHP with sets $\mathcal{N}_r$ and
$\{D'_i, i\in\mathcal{M}\}$.
\IF {OCMHP is infeasible}
\STATE Output ``Infeasible.'' Stop.
\ELSE
\STATE For $i\in\mathcal{M}$, let $\mathcal{S}'_i$ be the newly cached video set in step 2,
$\mathcal{S}_i=\mathcal{S}_i \cup \mathcal{S}_i'$,
$D_i'=D'_i - \sum_{k \in  \mathcal{S}'_i} s_k $.
\ENDIF
\STATE [Step 3]:
\FOR {$i \in \mathcal{M}$}
\STATE
 $\mathcal{N}_i=\mathcal{N}\setminus\mathcal{S}_i$
 is the set of videos not cached in node $i$.\\
 Solve Knapsack($\mathcal{N}_i$,$D_i'$,$i$), output $\mathcal{S}_i''$ \label{alg:line:knapsack}
\STATE  $\mathcal{S}_i=\mathcal{S}_i \cup \mathcal{S}_i''$.
\ENDFOR

\STATE [Step 4]: Calculate the objective value $H(\alpha)$
for solution $\{S_i, i \in \mathcal{M}\}$.
    Output $H(\alpha)$ and \{$\mathcal{S}_i, i \in \mathcal{M}$\}.
\end{algorithmic}
\end{algorithm}
%

At step 1,
we allocate storage
for the most popular videos on each serving node
using $\alpha$ of the total capacity.
We attempt to pack videos in each serving node with the objective of maximizing the
total hit ratio, such that
no more than $\alpha$ fraction of the total disk capacity  is used.
The problem is formulated as  the following
%
Reservation Packing problem ($\mathcal{N},\alpha$, $D_i, i\in\mathcal{M}$):
\begin{eqnarray}
&\max\ & \sum_{i\in\mathcal{M}}\sum_{k\in\mathcal{N}}s_k \lambda^k_i  y^k_i \label{eq:reservep} \cr
& \textrm{subject to} &
\sum_{i\in\mathcal{M}}\sum_{k\in\mathcal{N}}s_ky^k_i \le \alpha\sum_{i\in\mathcal{M}}D_i
\end{eqnarray}
and (\ref{eq:limcp-c1})(\ref{eq:limcp-c7}).
At step 2, we cache the  videos that were not cached at step 1
using the remaining disk capacity.
To ensure full coverage of all videos,
it is sufficient to maintain one copy
of these videos.
Thus, we change the constraint  in Eq. (\ref{eq:limcp-c4}) into the following equations:
   \begin{sequation}\label{equal:existequal}
    \sum_{i\in\mathcal{M}}y^k_i = 1, \forall k\in\mathcal{N}_r
  \end{sequation}
  where  $\mathcal{N}_r \subseteq \mathcal{N}$ denotes the set of less popular
  videos not cached in the first step.
  Additionally, we slightly modify the constraint in Eq. (\ref{eq:limcp-c1})
  to an equivalent constraint
  as follows:
  \begin{sequation}\label{equal:leftcapacity}
  \sum_{k\in\mathcal{N}_r}y^k_i s_k \le D'_i,  \forall  i\in\mathcal{M}
  \end{sequation}
%
We call the problem of maximizing (\ref{eq:shmpp}) subject to
(\ref{eq:limcp-c7})(\ref{equal:existequal})(\ref{equal:leftcapacity})
``One Copy Maximum Hit Problem (OCMHP).''

At step 3, we make use of the remaining space at each node
to further increase the hit ratio, and
formulate the problem as:
\begin{eqnarray}\label{eq:knapsack}
     &\max&
     \sum_{k\in\mathcal{S}''_i \subseteq \mathcal{N}_i}\lambda^k_i s_k \cr
&\textrm{s.t.}&
\sum_{k\in\mathcal{S}''_i\subseteq \mathcal{N}_i} s_k \le D'_i
\label{eq:oneknap}
\end{eqnarray}
at every serving node $i$, where $\mathcal{S}''_i$ is the variable to optimize.
Problem (\ref{eq:oneknap}) is
a typical Knapsack problem.

Finally, at step 4, we compute the objective value and output the cache
allocation.

Three steps remain for solving {MHP}:
(i) solving the Reservation Packing problem;
(ii) solving OCMHP;
(iii) solving the Knapsack problem (\ref{eq:oneknap}).
We discuss these steps in order.
%
\subsubsection{Solving the Reservation Packing problem}

Although the Reservation Packing problem  can be solved optimally
using dynamic programming,
it is probably too computationally expensive as our problem scale can be very
large.
Instead, we employ a greedy algorithm to solve the problem
and outline the procedure in Algorithm \ref{alg:grpreservation}.
(We will soon show that the greedy algorithm achieves
near optimal performance.)
At each iteration, we find the most popular pair $(i,k)$
among all feasible pairs and cache video $k$ at the node $i$.
A pair $(i,k)$ is feasible if the size of video $k$ is within
the remaining system capacity
as well as the remaining capacity on the node $i$.

\begin{algorithm}[h]
\caption {Greedy Algorithm for Reservation Packing}
\label{alg:grpreservation}
\begin{algorithmic}[1]
\STATE Initialize $\mathcal{S}_i=\emptyset$ for $i\in\mathcal{M}$,
$D=\alpha\sum_{i\in\mathcal{M}}D_i$,
$W=\{(i,k)|k\in\mathcal{N},i\in\mathcal{M}\}$.
\WHILE{$D>0$ and $W\ne \emptyset$}

\STATE $(i^*,k^*)=\arg \max_{(i,k)\in W} \lambda^k_{i}$
\IF{$D\ge s_{k^*}$ and $D_{i^*}'\ge s_{k^*}$} \label{line:sizecondition}
\STATE	
$\mathcal{S}_{i^*}$=$\mathcal{S}_{i^*}\cup\{k^*\}$,
$D=D-s_{k^*}$,
$D'_{i^*}=D'_{i^*}-s_{k^*}$
\ENDIF
\STATE $W=W\setminus \{(i^*,k^*)\}$
\ENDWHILE
\STATE Output $\mathcal{S}_i$ for all $ i \in \mathcal{M}$ and $\mathcal{N}_0=\cup_{i\in\mathcal{M}}\mathcal{S}_i$
\end{algorithmic}
\end{algorithm}

In a typical scenario, any individual video
size is much smaller than the disk capacity of the serving node.
We will show that under such a condition, the greedy algorithm
in Alg. \ref{alg:grpreservation} achieves near optimal performance.
The proof is omitted due to space limit and can be found in \cite{junhe2012a}.

\begin{theorem} \label{theorem:smhpp-ap}
If for all $i\in \mathcal{M},k\in \mathcal{N}$, $s_k\le \epsilon D_i$,
Alg. \ref{alg:grpreservation} is
at least $(1-\epsilon)(1-\frac{\epsilon}{\alpha M})$-suboptimal
for Reservation Packing Problem with $\alpha >0$, $\epsilon>0$,
where $M$ is the number of caching nodes.
\end{theorem}

\subsubsection{Solving OCMHP}
  OCMHP is a special case of the generalized assignment problem (GAP)\cite{Cattrysse1992}
where the sizes of items do not vary with the placement.
  GAP is a classical problem in combinatorial optimization, 
 which is proven to be NP-hard and even APX-hard to be approximated.
 Actually, the proof in Theorem \ref{theorem:smhpp-np}
also applies to complexity analysis for {OCMHP}.
Therefore, {OCMHP} problem is also strongly NP-hard.

The main purpose of this step lies in
caching all the videos in $\mathcal{N}_r \triangleq \mathcal{N} \backslash \mathcal{N}_0$,
rather than maximizing the total profit,
%
so we adopt the greedy method in \cite{romeijn2000class} 
with the weight function set to $\lambda_i^k$ in our implementation.
The details are omitted due to space limit.

%

\subsubsection{Solving the Knapsack problem}

  The problem formulated in (\ref{eq:knapsack}) is
 a classical {0-1 knapsack problem}, which is also NP-hard \cite{Martello1990}.
 Many algorithms for this problem can be found in \cite{Martello1990}.
 In our work,
 we adopt a greedy solution similar to Alg. \ref{alg:grpreservation} to obtain
 a sub-optimal solution.

\noindent
{\bf Complexity of $\alpha$-MHP Algorithm:}
The complexity of $\alpha$-MHP Algorithm depends on each step of
the algorithm.
To implement Alg. 2, we first sort the pairs $(i,k)$ in $W$ by $\lambda^k_i$,
then we go through all the pairs to complete the reservation packing.
Thus, the complexity of Alg. 2 is $O(MN\log(MN))$.
Similarly, the greedy algorithm for problem OCMHP  takes
time $O(N_r M \log(M) + N_r^2)$,
where $N_r$ denotes the size of the video set $\mathcal{N}_r$
that has not been cached in the previous step.
The complexity of step 3 is $O(MN \log N)$.
In summary, the total complexity of $\alpha$-MHP is
$O(MN\log(MN) + N^2)$.

\subsection{Finding Optimal $\alpha$}\label{sec:findalpha}

For a given problem instance,
the objective value $H(\alpha)$ produced by Alg. $\alpha$-MHP
is a function of $\alpha$.
What remains is to find the $\alpha$ that maximizes the objective
value $H(\alpha)$.
In general,  choosing a larger $\alpha$  increases the system utility
but decreases the chance of finding a feasible solution
to Problem OCMHP (as well as MHP), and vice versa.

\begin{figure}
\vspace{-10pt}
\setlength{\floatsep}{0pt}
\setlength{\abovecaptionskip}{-2pt}
\setlength{\belowcaptionskip}{-5pt}
 \centering
 \includegraphics[width=2.4in, height=1.8in]{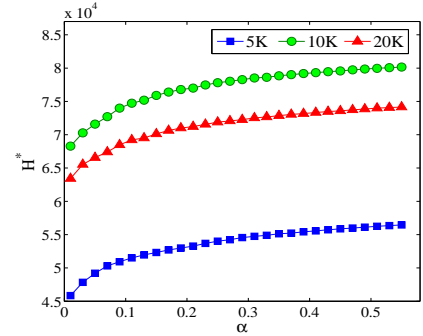}
\caption{Case Study for $H(\alpha)$}
\label{fig:case}
\end{figure}

We further investigate the property of
function $H(\alpha)$ by case studies.
We study a system consisting of 23 serving nodes.
Three instances are simulated with video library size of 5K, 10K, 20K,
respectively.
For each instance, we run Alg. $\alpha$-MHP
with $\alpha$ varying from 0 to 1 with step size of $0.01$.
We output $H(\alpha)$ found for each $\alpha$ in Fig. \ref{fig:case},
which shows that  $H(\alpha)$
produced by Alg. $\alpha$-MHP 
is an increasing function of $\alpha$ until  a feasible solution cannot be
found.
This confirms our intuition that
the more capacity is reserved for most frequently requested videos in each serving
node, the better objective value  can be found, until problem OCMHP becomes
infeasible.
Therefore, we apply binary search to find the optimal $\alpha$
in the interval [0, 1].
The main procedure, called System capacity Reservation Strategy (SRS), is summarized in Alg. \ref{alg:main}.

\begin{algorithm}
\caption {Main procedure - SRS}
\label{alg:main}
\begin{algorithmic}[1]
\STATE
Set the lower bound $\alpha_l = 0$ and compute $H(\alpha_l)$ using Algorithm
$\alpha$-MHP. If it returns ``Infeasible'',
we stop with the claim that the original {MHP} is infeasible.
\STATE Set the upper bound  $\alpha_u =1$ and compute
$H(\alpha_u)$ using Algorithm $\alpha$-MHP.
If $\mathcal{N}_r = {\O}$ at step 2 of $\alpha$-MHP,
stop and output this solution.
\STATE Otherwise, do binary search for $\alpha$ between
    $\alpha_l$ and $\alpha_u$ to find the maximum total utility $H(\alpha)$.
\end{algorithmic}
\end{algorithm}

%

\section{Solution to MRCP}
 In this section, we develop both centralized and distributed algorithms
 to solve the MRCP problem.
The centralized algorithm is guaranteed to be $\epsilon$-suboptimal while
the distributed scheme, which we refer to as LinkShare, provides traffic-aware and fine-grained control on the
source selection, which can be updated at sub-second levels.
Both of our schemes assume that the content placement is completed as a separate
step using the solution to MHP.

\subsection{Centralized Algorithm for MRCP}
\label{sec:subsolutiontomctp}

By aggregating the cost for all source-destination pairs on each link,
we can rewrite the objective (\ref{eq:mrcp}) of MRCP as:
   \begin{equation}\label{equal:mdtp2}
	\min\ g({\rm{\bf x}}) \triangleq
    \sum_{l\in\mathcal{L}}f_l^{ss}\zeta_l(f_l)
  \end{equation}
   where  ${\rm{\bf x}}$ is the vector containing all variables
   $\{x^k_{ji}\}$ and is implicitly contained in $f_l$ and $f^{ss}_l$.
   Together with Eqs. (\ref{eq:fssl}) and (\ref{eq:fl}),
   we can see that $g({\rm\bf x})$ is a convex function since $\zeta(f_l)$
   is convex.
   Therefore, we can solve it via convex optimization techniques.
   In this work, we adopt the interior-point method
   using the logarithmic function as the barrier \cite{Boyd2004}.
%
%
%
%
   For notational convenience, we write $f_l$ in (\ref{eq:fl}) as
   $f_l({\rm{\bf x}})$, $l=1,2\dots L$ 
 and define the barrier function:
     \begin{sequation}\label{equal:barrier}
      \phi({\rm{\bf x}})=-\sum_{l\in\mathcal{L}}\log(C_l-f_l(\rm{\bf x}))
    \end{sequation}
We then introduce a multiplier $m$ and consider the following problem:
\begin{sequation}\label{equal:interior}
\min\ mg({\rm{\bf x}})+\phi({\rm{\bf x}})
\end{sequation}
subject to  (\ref{eq:reachable})(\ref{eq:limcp-c6}).
Applying the duality analysis in \cite{Boyd2004},
we conclude that the optimal solution to (\ref{equal:interior})
is no more than $|\mathcal{L}|/m$-suboptimal,
provided that (\ref{eq:mrcp}) is feasible.
Consequently, we can obtain a solution which is guaranteed to be at most
$\epsilon$-suboptimal by taking $m \ge |\mathcal{L}|/\epsilon$ and  solving
problem (\ref{equal:interior}).
Standard interior-point method starts with a small $m$ and
sequentially solves the problem (\ref{equal:interior})
with increasing $m$. The detailed method is presented in
\cite{junhe2012a}
and is omitted here.

We note that as a preliminary step for solving the problem (\ref{equal:interior}),
we need to solve the feasibility problem, which turns out to be the min-max
link utilization problem solved in \cite{haiyongxie2012tecc}.

\subsection{Distributed Algorithm LinkShare for {MRCP}}\label{sec:solutiontomctp}
  In this subsection, we propose LinkShare,
  a distributed algorithm to MRCP, where
each serving node performs source selection independently in each round of time
duration $\Delta t$.
In order to minimize the total cost for the requests in the current
    round,
    we schedule the requests collaboratively to the sources with
    minimum cost at each serving node.
   We assume that the traffic information of each link is reported periodically
   to all serving nodes by the routers \cite{jiang2009cooperative}.
   To estimate the link loading
   between two reporting epochs,
   each node maintains a local loading table of all links independently.
   The local loading tables are updated either after a new local
   request is scheduled or the periodic reports are received.
   For each node $i$, we solve the
   problem:
\begin{eqnarray}
\label{equal:mdtpdistrlink}
        &\min\ & \sum_{k\in R_i}\sum_{j\in T_k} d^k_{ji} x^k_{ji} r_k\cr
        &\textrm{s. t.} &
    \sum_{j\in T_k}x^k_{ji} =1, \forall k\in R_i
\end{eqnarray}
and Eq. (\ref{eq:limcp-c6}).
To further reduce the complexity of the problem (\ref{equal:mdtpdistrlink}),
within each node $i$, we sequentially schedule each request and
update the local flow table
once after a request is scheduled. For each request $k$, we solve the problem:
\begin{eqnarray}\label{equal:mdtpdistrlink2}
        &\min\ & \sum_{j\in T_k} d^k_{ji} x^k_{ji} \cr
        &\textrm{s. t.} & \sum_{j\in T_k}x^k_{ji} =1
\end{eqnarray}
and Eq. (\ref{eq:limcp-c6}).
Problem (\ref{equal:mdtpdistrlink2}) can be solved analytically by finding the least-cost source,
i.e. $j^*=\arg \min_{j\in T_k} d'^k_{ji}$, where $d'^k_{ji}$ is temporary update of $d^k_{ji}$, assuming rate $r_k$ is added to the path $P(j,i)$.

We observe that most of the videos that need to be requested
from other serving nodes are of less popularity,
and typically have a small number of source nodes containing them.
The optimization process for problem (\ref{equal:mdtpdistrlink2}) works better
with more source nodes for a requested video.
Therefore, we sort the requested videos in the increasing order of the
number of source nodes containing them and then fulfill the video requests
in this order.
We list the resulting algorithm in Algorithm \ref{alg:distrmdtp}.

\begin{algorithm}
\caption { LinkShare for MRCP}
\label{alg:distrmdtp}
\begin{algorithmic}[1]
\MYREPEAT{every $\triangle t$ at each node $i\in\mathcal{M}$:}
\STATE Sort all requested videos in $R_i$ in the increasing order of  $|T_k|$.
\FOR {$k\in R_i$ }
\STATE Solve (\ref{equal:mdtpdistrlink2}) by finding the least-cost source $j^*$.
\STATE Request video $k$ from $j^*$.
\FOR {$l\in P(j^*,i)$}
\STATE Update local flow table, $f_l=f_l+r_k$
\ENDFOR
\ENDFOR
\MYUNTIL
\end{algorithmic}
\end{algorithm}

\subsection{Implementation issues}\label{sec:impissue}

We address some implementation issues that may arise in practical systems.

\subsubsection{Cost Functions}\label{ssubsection:cf}

One option for the cost function $\{\zeta_l(f_l)\}$
is to use a constant value independent of the link
loading, as used in \cite{applegate2010optimal}.
Ideally, we want the cost function to reflect
the congestion level of the links, so that the flows will avoid congested links.
A common option that meets this requirement is
to use the average delay in an M/M/1 queue, expressed by:
        $\zeta_l(f_l)= \frac{1}{C_l-f_{l}},\ f_{l}<C_l$.
To avoid the singular point at $f_l=C_l$,
we use the linear approximation for $f_l>\gamma C_l$,
where $0\le \gamma \le 1$,
as suggested in \cite{jiang2009cooperative}.
Precisely, we use the following expression as the cost function,
\begin{sequation}\label{eq:zeta}
        \zeta_l(f_l)= \begin{cases} \frac{1}{C_l-f_{l}}\ & \textrm{if } f_{l}< \gamma C_l, \cr
                      \frac{1}{(1-\gamma) C_l} + \frac{f_l - \gamma
                      C_l}{(1-\gamma)^2 c_l^2} & \textrm{otherwise}
                      \end{cases}
\end{sequation}
where $\gamma = 0.99$.
For such an option, the objective function in (\ref{eq:zeta}) is convex and
continuously differentiable.

\subsubsection{Congestion Avoidance}\label{ssubsection:ca}
Over-congestion causes significant delay of the traffic and sometimes
can result in packet losses if the buffer size is not sufficiently large.
%
To avoid over-congestion, we reserve a small fraction $\delta $ of the capacity
of each link $l$.
A source $j$ is unavailable to node $i$, if the aggregate flow $f_l$
on any link $l$ along the path $P(j,i)$ exceeds the threshold $(1-\delta) C_l$.
As a result, some requests may not be fulfilled
to avoid the congestion in the network.
Congestion-avoidance is an optional step in our scheme.


\subsubsection{Videos with long-duration}
\label{sec:ssbvd}
In practice, videos have different durations. A long-lasting video has several
issues  compared to a short video. First, a long-lasting video demands higher
bandwidth as it occupies the links for a long time. Second, some users may stop
watching the video before it finishes. To address these issues, we break long
videos into shorter ones, each having a fixed duration. Different pieces of an
original video have their own flow request frequency and may be requested and
routed independently.

\section {Performance Evaluation}\label{sec:numerical}

\subsection {Performance of the content placement algorithm}
\label{subsction:pnr}

The basic setup of our simulation is a network with 23 serving nodes and
$20,000$ video clips with size randomly and uniformly generated from 20MB to 400MB.
We control the capacity ratio, i.e. the ratio of the aggregate size of videos to
the aggregate capacity of nodes, to be between 0.2 and 0.8.
The requesting frequency for each video on each node is generated
based on the characteristic of the video and that of the node.
We first assign an integer value to each node as the
population parameter, denoting the number of users served by the node.
The population parameter is randomly drawn from a range, which is termed as
``population diversity'' henceforth.
For example, if the population diversity is $20 \sim 30$, it means the
population parameter is an integer randomly generated from [20, 30].
We then generate a Zipf distribution for all videos on each node,
with the exponent randomly selected within $0.7 \sim 0.9$.
In order to simulate diverse video distributions,
the ranks of videos are randomly permuted in every node.
The requesting frequency $\lambda^k_i$,
is set to the product of the Zipf factor for video $k$ on the node $i$ and
the population parameter of the node $i$.

To evaluate the performance of Alg. \ref{alg:main} (denoted as SRS),
we compare it with
the method suggested in \cite{haiyongxie2012tecc},
where each serving node independently keeps a uniform $\alpha$ fraction
of its storage capacity for most frequently requested videos, and the rest
of the capacity is devoted to covering all remaining videos collaboratively.
We find the optimal $\alpha$ by enumerating all possible $\alpha$ with
precision 0.01.
We name it ``individual reservation strategy'' or IRS for short.
Additionally, we derive an upper bound of the solution
by relaxing the binary constraint (\ref{eq:limcp-c7})
to be $y^k_i\in [0,1]$,
and solving the resulting linear programming (LP) problem for the MHP problem.

\begin{figure*}
\setlength{\belowcaptionskip}{-10pt}
\centering
\begin{minipage}[tb]{0.3\linewidth}
   \includegraphics[width=2.3in,height=2.0in]{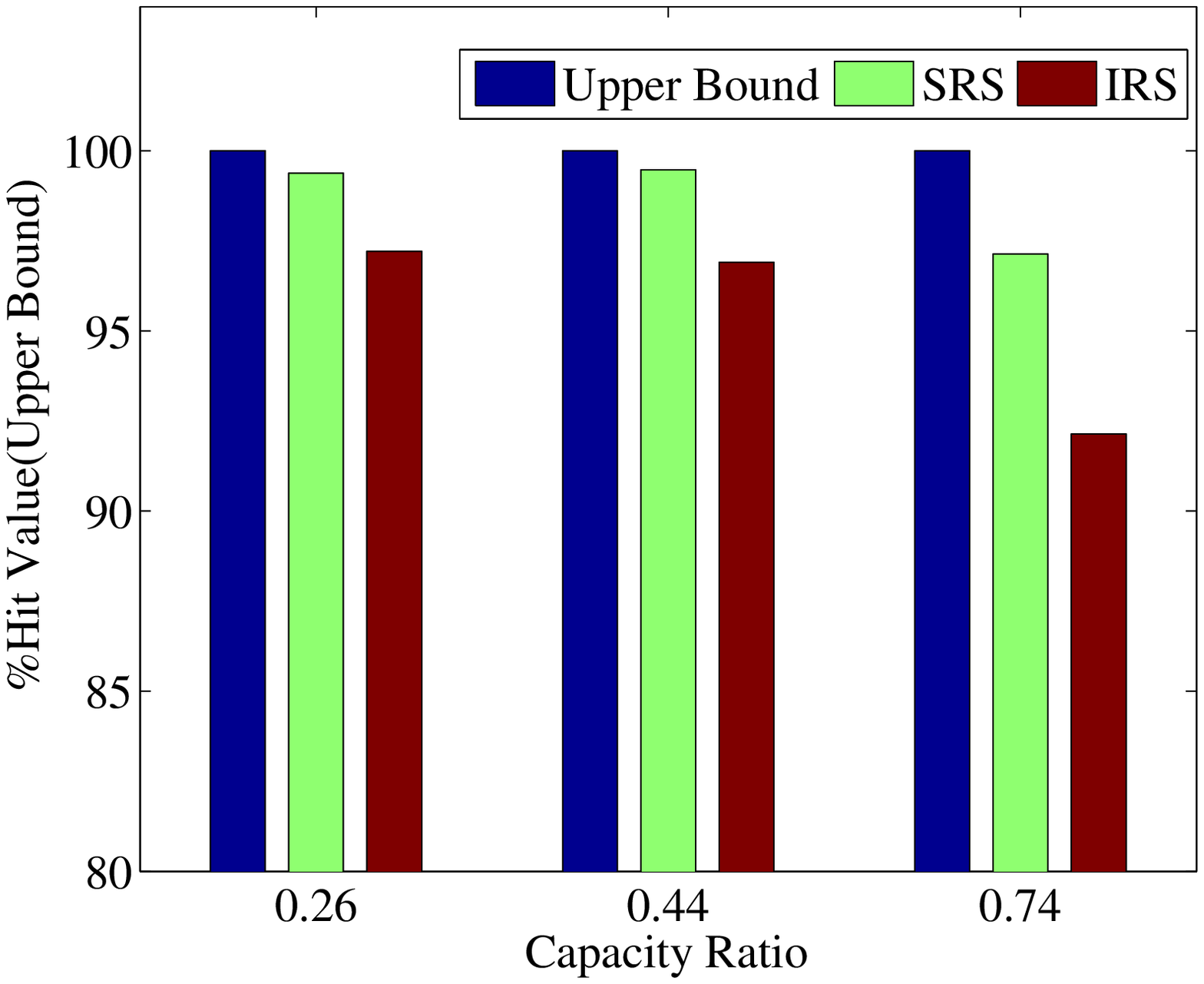}
   \caption{Hit ratio v.s. Capacity ratio}
   \label{fig:cr}
  \end{minipage}
\begin{minipage}[tb]{0.3\linewidth}
    \includegraphics[width=2.3in,height=2.0in]{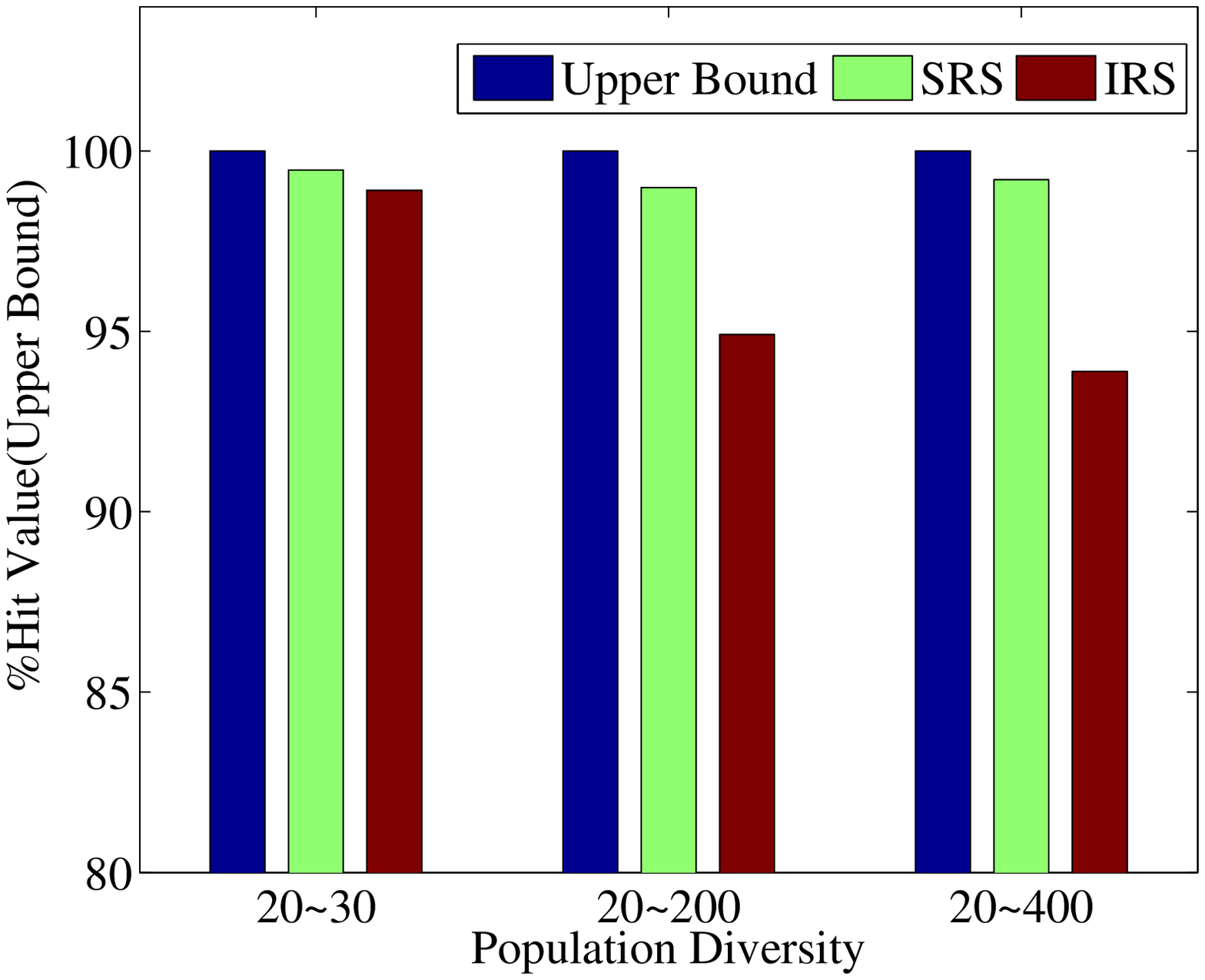}
   \caption{Hit ratio v.s. Node Diversity}
   \label{fig:dn}
\end{minipage}
\begin{minipage}[tb]{0.3\linewidth}
\setlength{\abovecaptionskip}{10pt}
    \includegraphics[width=2.1in,height=1.82in]{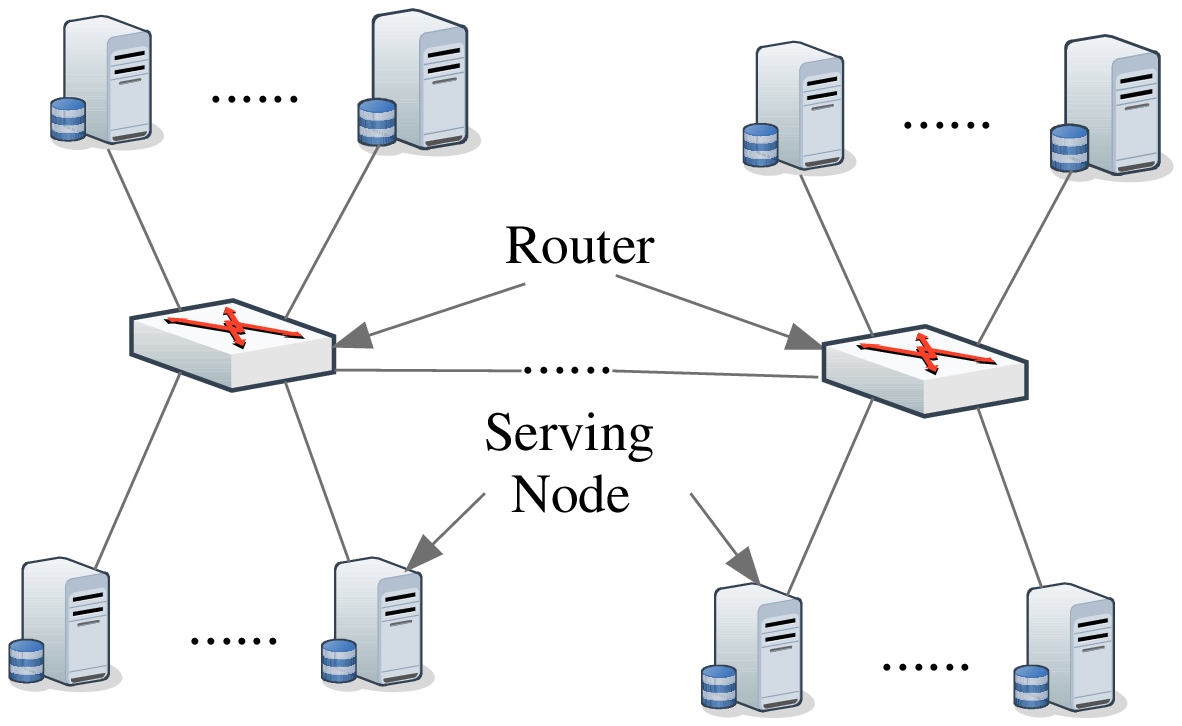}
   \caption{System Topology}
   \label{fig:star}
\end{minipage}
\end{figure*}

In Fig.  \ref{fig:cr}, we show the hit ratio vs.
the capacity ratio, where the population diversity is $20\sim 30$ and the
capacity ratio is around $0.26$, $0.44$, $0.74$ respectively.
Fig. \ref{fig:dn} compares the performance
with different population diversity
under a fixed capacity ratio of $0.44$.
From these two figures, we can see that
our {SRS} algorithm is always better than {IRS},
and its performance is typically within $1\%\sim 3\%$
of the upper bound obtained by linear relaxation.
We also notice that the performance of IRS
is rather sensitive to
the population diversity and the capacity ratio,
while that of SRS is quite stable.

We also evaluate the running time of our algorithm when
solving a larger instance consisting of 56 serving nodes
with caching capacity varying from 1.2TB to 2.4TB,
and 200,000 video clips with sizes randomly generated from 20MB to 400MB.
The capacity ratio is $0.46$ and the population diversity is $20\sim 30$.
It takes 1774 seconds and 1.8GB memory to find a SRS solution
with precision of $0.005$ for $\alpha$.
The result is $98.55\%$ of the upper bound obtained by linear relaxation.
All the above experiments are run on a server with 3.20GHz
Intel Xeon processor and 64GB of memory.

\subsection {Performance of the Source Selection algorithm}
We use the system with 56 serving nodes metioned
in section \ref{subsction:pnr} to
evaluate our algorithms for the source selection problem.
We simulate a mobile core network with
8 routers connected via links of 10Gbps and
7 serving nodes (i.e., serving gateway) attached to each router via links of 1Gbps.
Fig. \ref{fig:star} shows the basic topology.
In our experiments, we use a uniform video rate of 128Kbps. We adopt the link cost model in section \ref{ssubsection:cf}.
Requests are randomly generated for each node in every slot according
to the frequency distribution $\{\lambda^k_i\}$.
A request of video $k$ at node $i$ is called {\em collaborative request} if
video $k$ is not found in the local cache of the node $i$.
The average frequency of collaborative requests, called {\em traffic intensity},
plays an important role in
determining in-network traffic, and thus is a controlling factor in our
experiments.
%
%

{\noindent \bf Reference algorithms:}
For comparison, we implement four reference algorithms.
\begin{icompact}
 \item Traffic Engineering Approach (TE): the source selection
 is determined based on the goal 
 of minimizing the maximum congestion level on all links,
  which was investigated in \cite{haiyongxie2012tecc}.

 \item End-to-End Approach (E2E)\footnote{The original E2E approach requires one end-to-end measurement for
 each source-destination pair of requests and
 is impractical in a VoD system with burst requests.
 In the later simulations, we apply it to our framework,
 with each source-destination delay measured at most once in a round.
 Thus, the aggregate number of measurements is bounded by $O(M^2)$ for a round.}:
 the server $j\in T_k$ with the least end-to-end
 latency (measured) to node $i$ is selected.
 This principle is applied in
    Akamai \cite{Su2006}.

\item Nearest-Source Approach (NS): the server $j\in T_k$ with the nearest
distance (measured in hops) to node $i$ is selected.
This approach is suggested and evaluated in \cite{applegate2010optimal}.

\item Random approach: the source server is randomly selected from $T_k$.
\end{icompact}
Since the performance of source selection is influenced by the instantaneous
link states as well as the instantaneous request patterns,
we next consider both static and dynamic scenarios to evaluate the above approaches.

\subsubsection{Static Scenario}

In a static scenario,
we run different solutions for one slot and compare the aggregate latency caused.
NS  works exactly in the same way as E2E in the one-slot simulation because the
initial link loading is set to be all equal.

At the beginning of the slot, each link is assumed to be $\frac{1}{4}$-full.
With traffic intensity over the range of 20 to 120, we evaluate the algorithms and
show the aggregate cost in fig. \ref{fig:oneslot}.


From Fig. \ref{fig:oneslot}, we find that the TE approach, which aims to
minimize the maximum link utilization, has the worst performance in terms of
the aggregate link cost, even worse than the Random scheme.
LinkShare algorithm performs slightly better than E2E model, and the centralized
algorithm performs the best.
Both the centralized algorithm and the TE approach require solving large-scale
linear programming problems and are not amenable for implementation in real-time
environments. Thus, they are not compared in the dynamic scenario below.

\subsubsection{Dynamic Scenario}

In the dynamic scenario, we compare the performance of four
distributed algorithms,
LinkShare, E2E, NS, and Random,
in a system with a continuous workload for $100$ slots (each slot has
duration 0.1s).
The traffic intensity is $160$.
Requests are re-scheduled every $10$ slots based on the arguments in section \ref{sec:ssbvd}.
Additionally, the states of links are reported to each node every other slot in
the LinkShare algorithm.
In the E2E model, we assume that accurate end-to-end latency can be measured by nodes.

\begin{figure*}
\setlength{\belowcaptionskip}{-10pt}
\centering
\begin{minipage}[h]{0.3\linewidth}
\centering
\includegraphics[width=2.3in, height=2.0in]{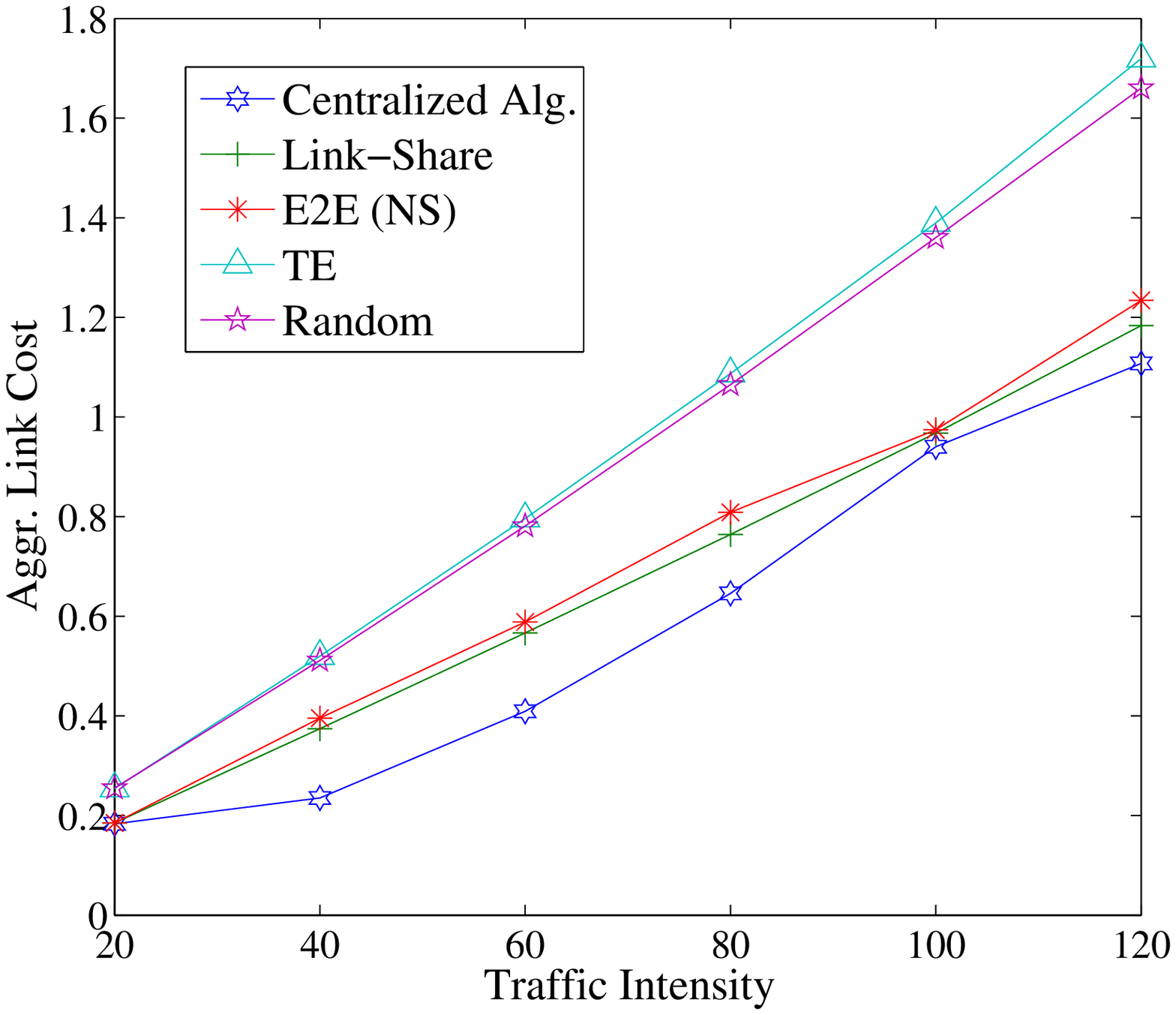}
\caption{Static Scenario}
\label{fig:oneslot}
\end{minipage}%
\begin{minipage}[h]{0.3\linewidth}
\centering
\includegraphics[width=2.3in, height=2.0in]{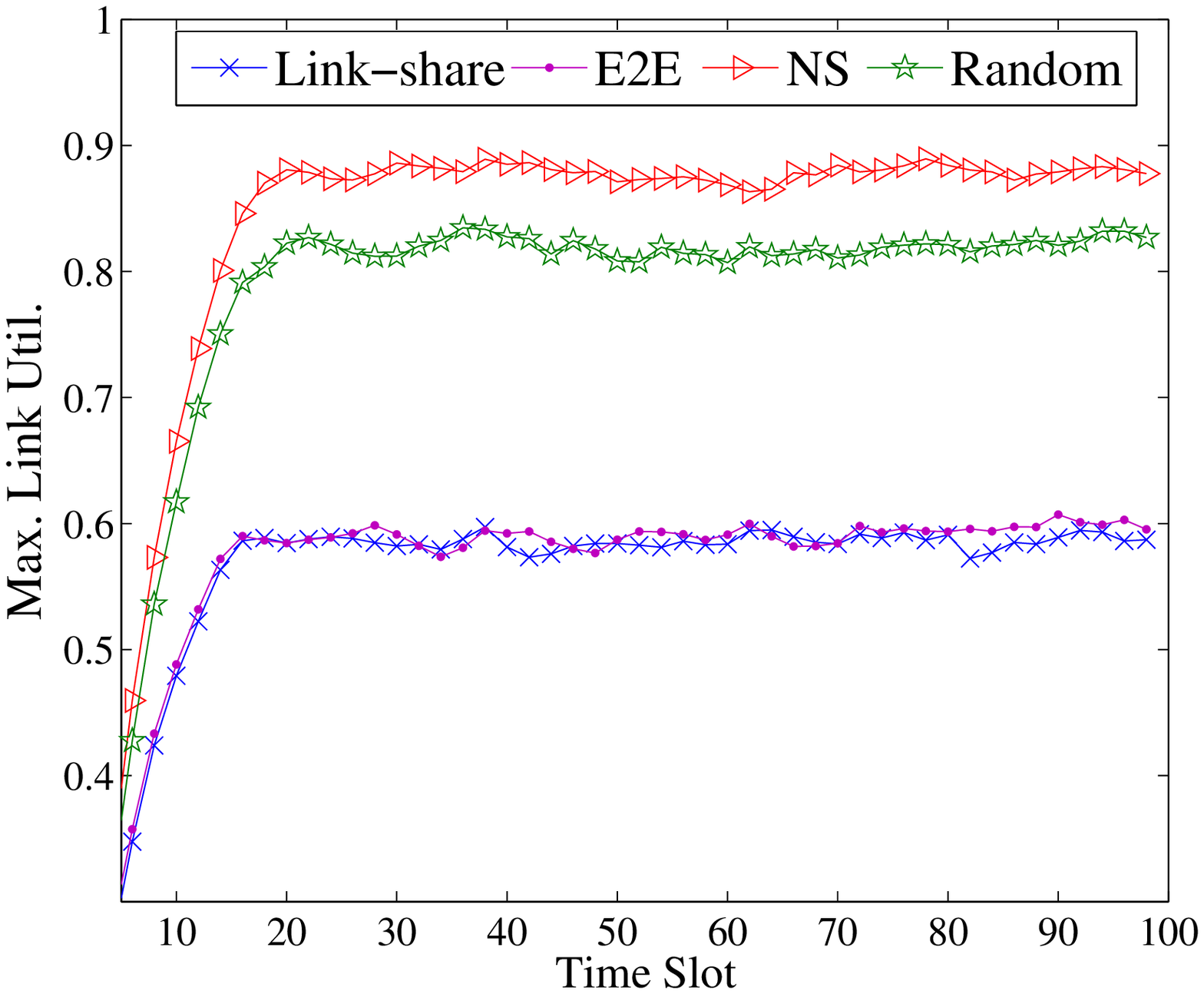}
\caption{TE Metric }
\label{fig:ste}
\end{minipage}
\begin{minipage}[h]{0.3\linewidth}
\centering
\includegraphics[width=2.3in, height=2.0in]{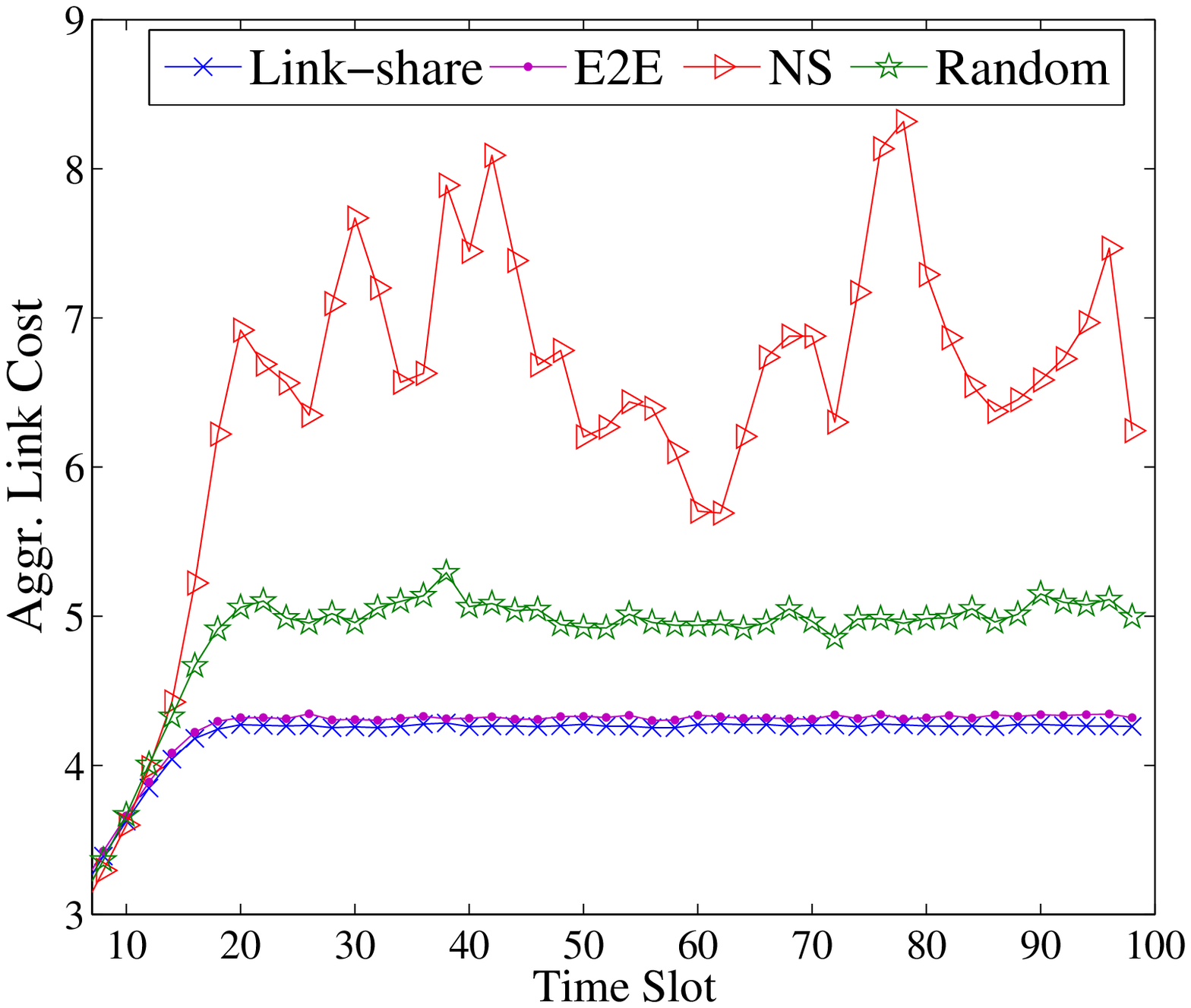}
\caption{Aggregate Cost }
\label{fig:scost}
\end{minipage}%
\end{figure*}

Fig. \ref{fig:ste} and Fig. \ref{fig:scost}
show the performance of these distributed algorithms under
the traffic-engineering metric, i.e. the maximum link utilization,
and the aggregate-link-cost metric respectively.
In addition, with continuous system load, the nearest-source strategy performs
the worst, even worse than the random strategy.

Both the E2E approach and our LinkShare method  perform very well.
The E2E approach relies on end-to-end measurement of the path latency
while our LinkShare method assumes the periodical state report from the routers.
So they can be applied to different conditions (depending on whether
the periodic state report is available  from the routers).
Later, we will show that E2E approach requires more network overhead
than LinkShare.


\begin{figure*}
\setlength{\belowcaptionskip}{-10pt}
\centering
\begin{minipage}[h]{0.3\linewidth}
\centering
\includegraphics[width=2.3in, height=2.0in]{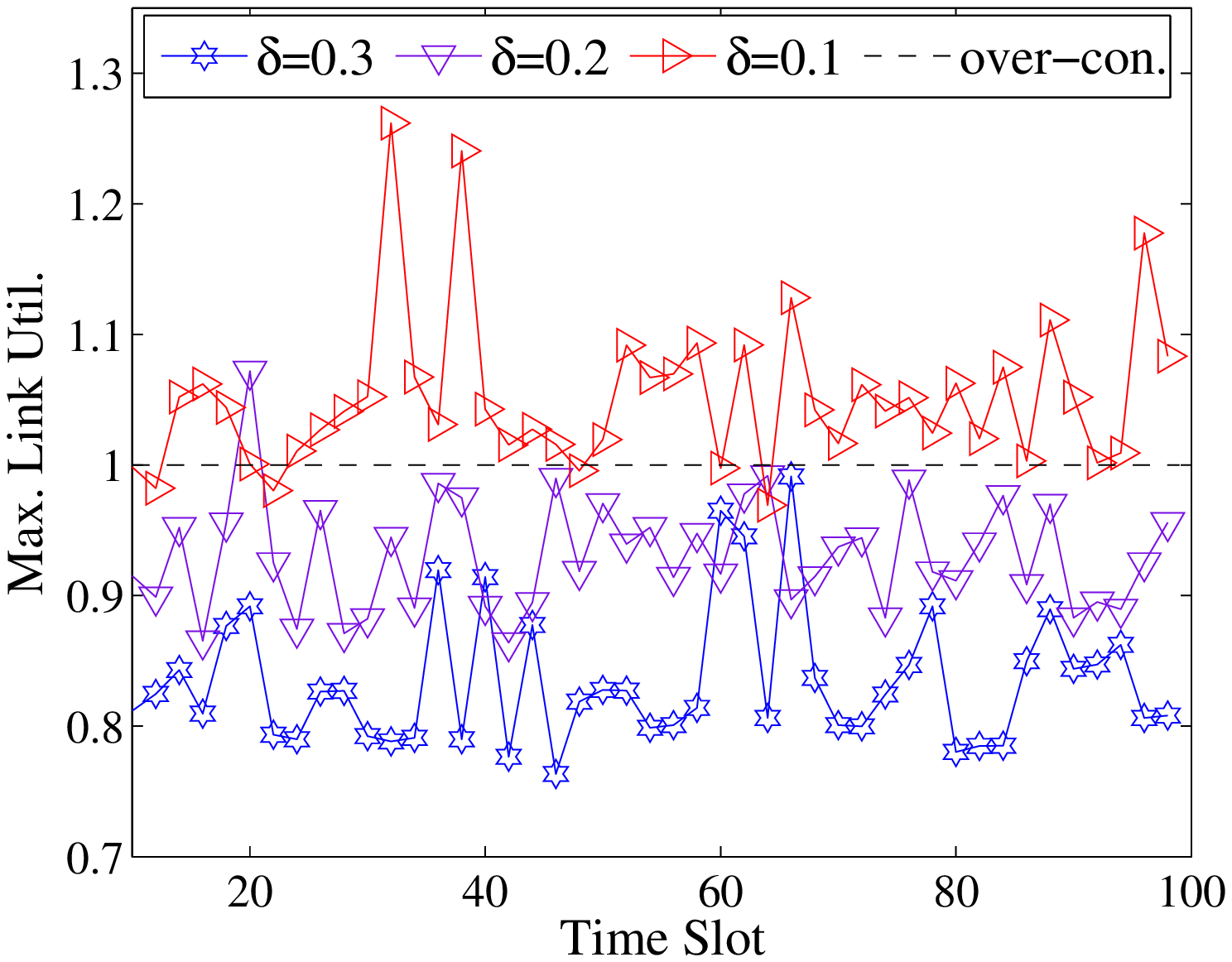}
\caption{Max. Link Utilization v.s. $\delta$ }
\label{fig:holdte}
\end{minipage}%
\begin{minipage}[h]{0.3\linewidth}
\centering
\includegraphics[width=2.3in, height=2.0in]{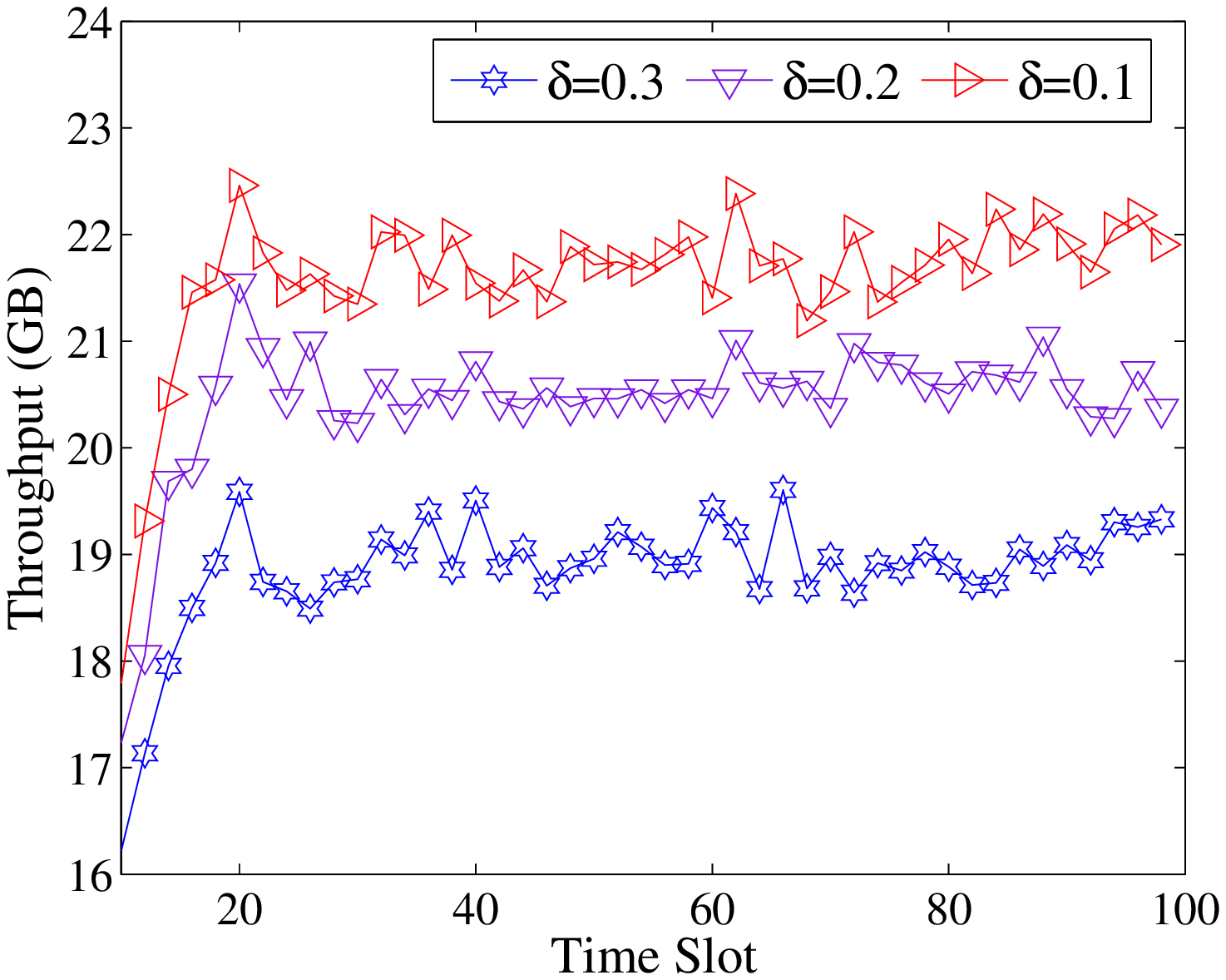}
\caption{ Throughput v.s. $\delta$}
\label{fig:holdth}
\end{minipage}
\begin{minipage}[h]{0.3\linewidth}
\centering
\includegraphics[width=2.3in, height=2.0in]{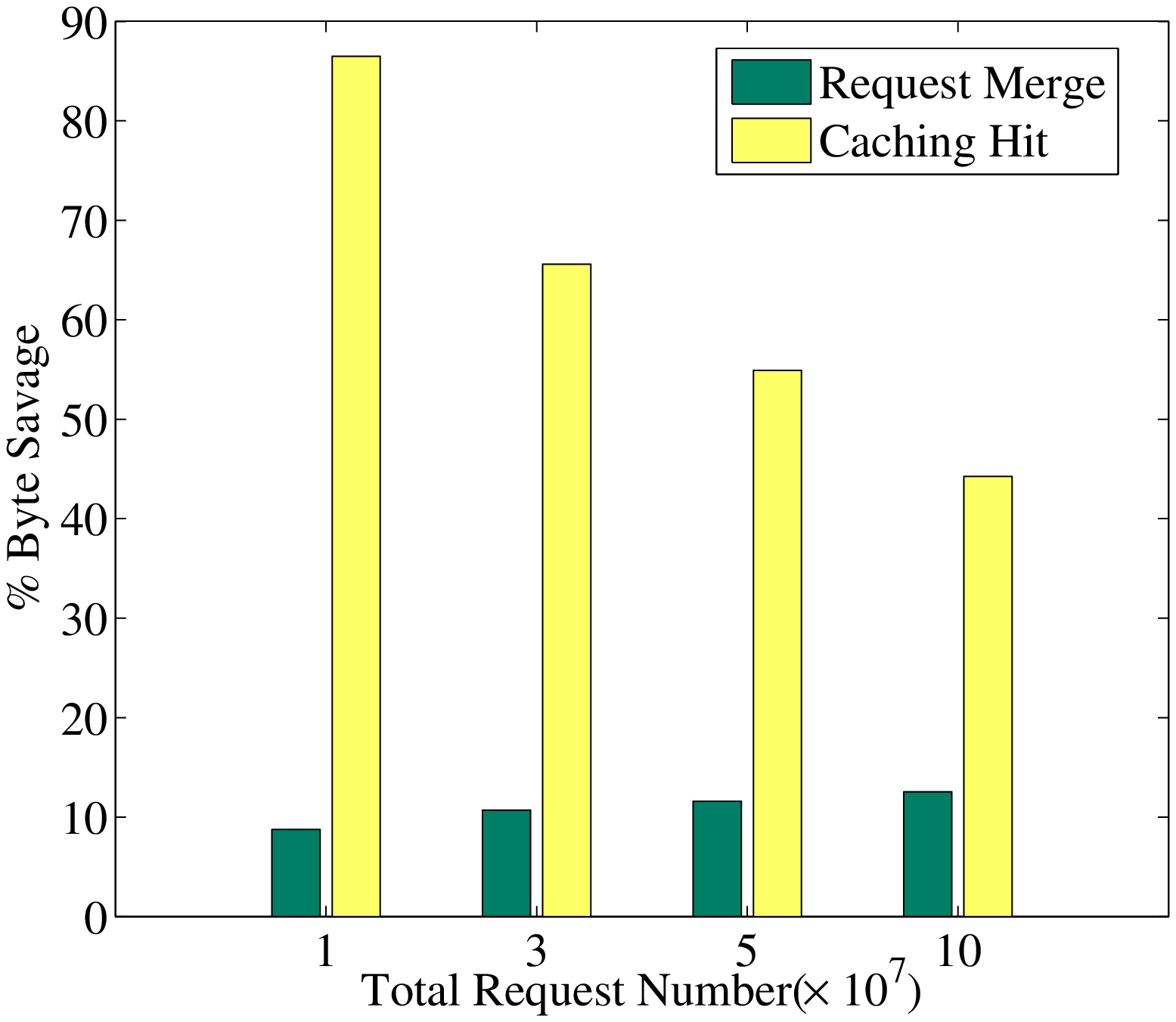}
\caption{Cache hit and merge saving}
\label{fig:hit}
\end{minipage}%
\end{figure*}

\subsubsection{Congestion Avoidance}
  As mentioned in section \ref{ssubsection:ca},
 our LinkShare approach can reserve a fraction $\delta$ of the link capacity
 to avoid network congestion caused by non-cooperative traffic generated from
 neighboring nodes.
%
With congestion avoidance, source $j$ is unavailable to node $i$,
if the path $P(j,i)$ contains links with flow amount
(read from local flow table)
exceeding $1-\delta$ fraction of the capacity.
Accordingly, video $k$ is unavailable to node $i$,
if $k \notin S_i$ and all sources in $T_k$ is unavailable to $i$.

 We load the system with heavy traffic intensity of $900$
 and evaluate the performance with $\delta=0.1, 0.2, 0.3$, respectively.
We then run the simulations for $100$ slots and
show the traffic engineering metrics and
the in-network throughput metrics in Fig. \ref{fig:holdte} and Fig. \ref{fig:holdth}.
As shown in the figures, the more capacity is reserved,
the less congestion the LinkShare approach produces,
although the in-network throughput also decreases.
In practice, we can find a good $\delta$ through detailed system-level
simulations.
For instance, in the network we simulated,  $\delta = 0.2$ appears to be
a good choice.

\subsubsection{Request merging and cache hit}
We also evaluate the efficiency of request merging and caching hit.
Fig. \ref{fig:hit} shows the traffic saving
in percent for both request merging and caching hit
with the total request number of $1,3,5,10(\times 10^7)$, respectively.
We find that more than $80\%$ of traffic can be saved by
caching hit when the system is under light load.
By contrast, during the peak time with heavy load,
 the probability of repeated requests during the same round increases.
As a result, nearly $10\%$ of traffic can be saved by request merging.

\subsubsection{Overhead Analysis}

Both
Link-share method and E2E method
require the network state information, thus incurring  some control overhead.
We provide an estimate of the total extra bandwidth on all links introduced by
these two methods.

In Link-share method, the overhead is produced by
the periodic link-state report from all routers to all serving nodes.
Each router can build a multicast tree to disseminate the link states to all
serving nodes.  Thus, it will need 63 (which is the number of
links) hops to reach all serving nodes in our simulated network in every
reporting cycle.
Each link-state reporting packet contains 32 bytes,
including a 4-byte payload of link load,
a 8-byte UDP header and a 20-byte IP header.
With 63 links in the simulated system,
the aggregate size of all reporting data over all links is
$63 \times 63 \times 32\approx 124$ KBytes.
If a reporting cycle has 2 slots and each slot is 0.1 second,
the aggregate overhead is about 4.84Mbps for the system.
Note that this is the total bandwidth introduced on all links in the system.

E2E approach relies on the end-to-end latency,
which is typically obtained by the ICMP (Internet Control Message Protocol)
echo request and echo reply messages.
As a result, the overhead of E2E approach
consists of the probing message between all pairs of the caching nodes.
Every ICMP echo packet has a default size of 32 bytes.
In our simulated system,
The total number of probes is $56*55=3080$ per slot.
The average hop length in our simulation is $5.5$ hops, resulting in an average round-trip length of $11$ hops.
Therefore, the aggregate overhead is $3080*11*32\approx 1.03$ MBytes per slot,
which is  about $82.71$Mbps for the whole system.

\section{Conclusion}\label{sec:conclusion}
 To reduce the network cost for VoD services
 in broadband mobile core networks,
 we propose a novel framework for collaborative in-network video caching in this paper.
 We formulate the caching problem as minimizing the total network cost while covering 
 a subset of the videos with high request frequency. 
 We decompose the problem into two subproblems: a collaborative content placement
 subproblem and a source selection subproblem.
We propose an efficient heuristic algorithm
for the content placement subproblem based on the long-term average video request
frequency.
With instantaneous information on the request patterns and link load,
we develop both centralized and distributed algorithms 
for the dynamic source selection subproblem. 
We also discuss several implementation issues in practical systems.
%
We perform extensive simulations to evaluate our proposed schemes.
 Simulation results show that our heuristic algorithm for the placement subproblem
 achieves solutions that are within $1-3\%$ of the optimal values,
 and our distributed algorithm Link-share is more efficient and requires less
 overhead  than existing algorithms.
 We also show  that up to $10\%$ of traffic can be saved by request merging,
 and up to $80\%$ can be saved
 by caching hit under light load of requests.



\bibliographystyle{unsrt}

\appendices
\section{Proof of Lemma 1}
\begin{proof}
 Without loss of generality, we assume that the items are sorted such that:
\begin{eqnarray}
\label{eq:gknap1}
 \frac{p_1}{a_1}\ge \frac{p_2}{a_2} \dots\ge \frac{p_{J}}{a_{J}}.
\end{eqnarray}
Let $\kappa$ be the index of the first item that is rejected by KnapsackGA,
$G$ be the maximum profit found by KnapsackGA and
$OPT$ be the optimal result for the Knapsack problem.
Then we have (i) $G \ge p_1+p_2+\dots +p_{\kappa-1}$,
(ii) $a_1+a_2+\dots +a_{\kappa} >B $,
and (iii) $p_1 + p_2 + \dots +p_{\kappa} \ge OPT$.
Thus, from Eq. (\ref{eq:gknap1}),
\begin{eqnarray}
\nonumber
 p_1+p_2+\dots+p_\kappa &\ge& (a_1+a_2+\dots+a_\kappa)p_\kappa/a_\kappa \\
\nonumber
\Rightarrow p_\kappa &\le& a_\kappa(p_1+p_2+\dots+p_\kappa)/B \\
\nonumber &\le& \varepsilon(p_1+p_2+\dots+p_\kappa)
\end{eqnarray}
where the last inequality holds because $a_{\kappa} \le \varepsilon B$.
Rearranging the above equation, we have:
$$
(1 -  \varepsilon) (p_1+p_2+\dots+p_\kappa) \le  p_1 + p_2 + \dots + p_{\kappa -1}
$$
Now we have:
\begin{eqnarray}
G &\ge& p_1 + p_2 + \dots + p_{\kappa -1} \cr
  &\ge&  (1 -  \varepsilon) (p_1+p_2+\dots+p_\kappa) \cr
  &\ge& (1-\varepsilon) OPT
\end{eqnarray}
%
\end{proof}

\section{Proof of Theorem 3}
\begin{proof}
 Provided that MHP is feasible for  $\alpha_2$,
let $\mathcal{N}_{0}^{\alpha_1}$ and $\mathcal{N}_{0}^{\alpha_2}$
be the cached video set
 after step 1 in Alg. $\alpha_1$-MHP and $\alpha_2$-MHP, respectively.
We note that
the first step of $\alpha_2$-{MHP} can be naturally divided into two phases.
In phase 1, we run Alg. \ref{alg:grpreservation} until
the remaining capacity is $1-\alpha_1$ and in phase 2,
we continue the algorithm until the remaining capacity is $1-\alpha_2$.

We now compare Step 2 in $\alpha_1$-MHP and phase 2 of Step 1 plus Step 2
in  $\alpha_2$-MHP. Both have the same capacity and cache the
same set of remaining videos $\mathcal{N} \backslash \mathcal{N}_0^{\alpha_1}$.
The former does not allow to duplicate videos cached in the system
while the latter does.
Thus, if $\alpha_2$-MHP can generate a  feasible solution,
so can $\alpha_1$-MHP, given that Alg. \ref{alg:grpreservation} is optimal.
  The converse part of the theorem can be proven by contradiction.
   \end{proof}

\end{document}